\documentclass[twocolumn,showpacs,preprintnumbers,amsmath,amssymb]{revtex4}

\usepackage{graphicx}
\usepackage{dcolumn}

\begin{document}

\title{Origin of the peak-dip-hump structure in the photoemission spectra\\ of Bi$_2$Sr$_2$CaCu$_2$O$_8$}
\author{A. A. Kordyuk$^{1,2}$, S. V. Borisenko$^1$, T. K. Kim$^1$, K. Nenkov$^1$, M. Knupfer$^1$, M. S. Golden$^1$, J. Fink$^1$, \\H. Berger$^3$, R. Follath$^4$}
\address{$^1$ Institute for Solid State and Materials Research, IFW Dresden, P.O.Box 270016, D-01171 Dresden, Germany}
\address{$^2$ Institute of Metal Physics of National Academy of Sciencies of Ukraine, 03142 Kyiv, Ukraine}
\address{$^3$ Institut de Physique Appliqu\'ee, Ecole Politechnique F\'ederale de Lausanne, CH-1015 Lausanne, Switzerland}
\address{$^4$ BESSY GmbH, Albert-Einstein-Strasse 15, 12489 Berlin, Germany}

\date{September 21, 2001}%

\begin{abstract}
The famous peak-dip-hump lineshape of the ($\pi$,0) photoemission spectrum of the bilayer Bi HTSC in the superconducting state is shown to be a superposition of spectral features originating from different electronic states which reside at different binding energies, but are each describable by essentially identical single-particle spectral functions. The 'superconducting' peak is due to the antibonding Cu-O-related band, while the hump is mainly formed by its bonding counterpart, with a $c$-axis bilayer coupling induced splitting of about 140 meV.
\end{abstract}


\maketitle

The low energy photoemission spectra from the ($\pi$,0)-point of the Brillouin zone (BZ) of the majority of the high temperature superconducting cuprates (HTSC) in the superconducting state exhibit the now famous 'peak-dip-hump' (PDH) lineshape \cite{Dessau}. This lineshape was, up to the present, widely believed to be the result of a single spectral function (see for instance Refs.~\onlinecite{DingPRL96,NormanPRL97,CampuzanoPRL99}) which is caused by, e.g., a strong coupling to bosons \cite{CampuzanoPRL99,LanzaraNature}, and the details of which are expected to reveal, at a fundamental level, the identity of the interactions involved in the generation and perpetuation of the superconducting state in these systems \cite{LanzaraNature,FengScience,Orenstein}.

In this Letter we report the results of a detailed, high resolution ARPES study of this important BZ region in overdoped, modulation-free Pb-Bi2212 crystals using a wide range of excitation energies. The central result is that the strongly differing photon energy dependencies of the intensities of the 'peak' and 'hump' in the ($\pi$,0) spectrum essentially exclude a scenario in which the peak, dip and hump features originate from a single-band spectral function. The data argue for a straightforward picture in which - at least for the overdoped compounds - the 'superconducting' peak in the ARPES spectra is related to the antibonding CuO-bilayer band the role of which in the origin of superconductivity in the bi- or trilayer cuprates should be reconsidered and the hump is mainly formed by the bonding band.

The ARPES experiments were carried out using angle-multiplexing photoemission spectrometers (SCIENTA SES200 and SES100). The momentum distribution maps and series of energy distribution curves (MDMs and EDCs) were measured at 300K or 39K using $h\nu=$21.218 eV photons from a He source, as described elsewhere \cite{BorisPRL,BorisPRB,KordXXX}. The ($\pi$,0) EDCs were recorded using radiation from the U125/1-PGM beamline \cite{Follath} at BESSY. The total energy resolution ranged from 8 meV (FWHM) at $h\nu=$ 17--25 eV to 22.5 meV at $h\nu=$ 65 eV, as determined for each excitation energy from the Fermi edge of polycrystalline gold (which also gives the energetic calibration in each case).
All data were collected on two similar overdoped single crystals of Pb-Bi2212 ($T_c$=69K), except for the Fermi surface map taken from pure Bi2212 ($T_c$=89K). All ($\pi$,0)-EDCs were measured at a temperature of 27K - deep in the superconducting state.
 
\begin{figure}[b!]
\includegraphics[width=8.47cm]{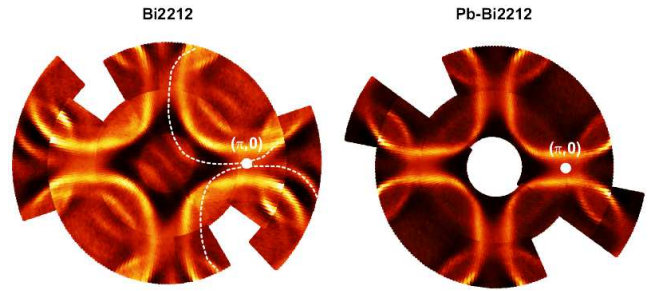}%
\caption{\label{Maps} Fermi surface maps of pure Bi2212 (left panel) and Pb-Bi2212 (right panel) measured at room temperatures.}
\end{figure}

We start by presenting the momentum distribution maps of the $E_F$ photoemission intensity for pristine and Pb-doped Bi2212 in Fig.~\ref{Maps}. When measured in a normal state as was done here, local intensity maxima in such datasets map out the Fermi surface (FS) \cite{BorisPRB}. However, it is well known that in the case of pure Bi2212 these intensity maps contain an additional set of features originating from the scattering of the outgoing photoelectrons on the incommensurably modulated BiO-layers \cite{BorisPRL,Fretwell}. Such diffraction replicas are extrinsic in nature and can be easily identified given MDMs which cover a sufficiently large portion of the {\bf k}-space. With the aid of Fig.~\ref{Maps}, it also easy to estimate to which extent the lineshape of a given EDC may be contaminated by these extrinsic replicas (see the white dashed lines). Obviously the most perilous region for examination would be one in which a high density of different features overlap or are closely separated. Thus as it is clear that the much-discussed PDH-spectrum from near the ($\pi$,0)-point of Bi2212 originates from exactly such a {\bf k}-space region, the crucial additional security of investigating the ($\pi$,0) PDH lineshape in modulation-free samples is apparent.

The use of the modulation-free samples had allowed us to clarify the FS topology \cite{BorisPRL} and show that contradictions in this point could be explained by the presence of the diffraction replicas and by strong influence of excitation energy dependent matrix elements on photoemission stpectra from the ($\pi$,0) region \cite{BorisPRB,SibyllePRB}. On the other hand, by analysing changes in the lineshape upon varying experimental parameters linked to the matrix elements, one gains insight into the nature of a given feature. For instance, the observation, that the ($\pi$,0) PDH lineshape turned out to be insensitive to an alteration of the polarisation conditions at a fixed photon energy  \cite{DingPRL96}, has been one of the cornerstones of many of the single-band theories developed to explain the PDH lineshape in the superconducting state. In the present work we utilize a possibility to significantly alterate the photoemission matrix elements by variation of the excitation energy. In this context it is interesting to note that the PDH spectrum reported in all intensively referenced publications over the last ten years has been recorded on the pristine Bi2212 and using only a very narrow range of photon energies (19--22.4 eV) \cite{Dessau,DingPRL96,NormanPRL97,CampuzanoPRL99,FengScience,Others}.

\begin{figure}[t!]
\includegraphics[width=8.47cm]{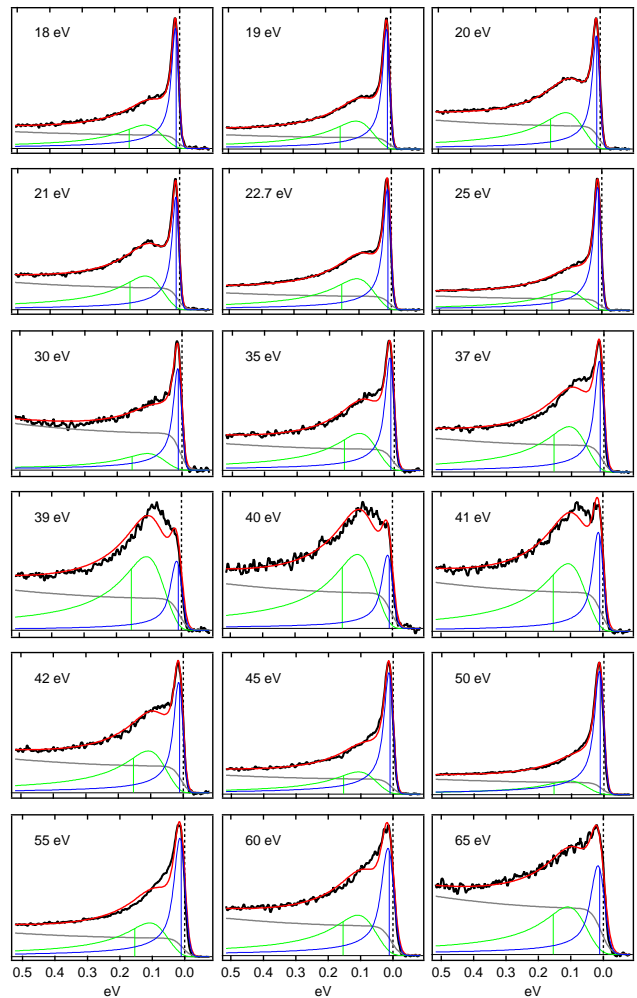}%
\caption{\label{Edep} The ($\pi$,0) photoemission spectra from the overdoped (69K) sample for different excitation energies: the black lines show the experimental data and the color lines represent the results of fitting procedure which is described in the text.}
\end{figure}

Fig.~\ref{Edep} shows a collection of superconducting state ($\pi$,0) photoemission spectra for overdoped Pb-Bi2212 recorded using different excitation energies $h\nu$ (18--65 eV). The black lines show the experimental data and the red lines represent the results of a fitting procedure which is described below. Upon a visual inspection of the raw experimental data in Fig.~\ref{Edep}, it is evident that the PDH lineshape can no longer be considered to be the result of single band spectral function with a sophisticated self-energy, as although at some excitation energies (e.g. 20 eV) all three components (peak, dip and hump) are present, there are also photon energies at which there is virtually no dip (25 or 42 eV), no hump (50 eV) or even no peak (39 eV).  A further, firm conclusion which can be made from inspection of the raw data alone, is that the hump itself cannot be considered as a feature appearing at a fixed binding energy for all $h\nu$ (compare the EDCs for $h\nu$ 20 and 37 eV, for example). For further analysis a fitting procedure has been applied.

We begin with the discussion of a two-peak fitting procedure, which is intuitively obvious bearing in mind the recently observed bilayer splitting \cite{FengPRL01,ChuangPRL01} in the Bi2212. The results are shown in Fig.~\ref{Edep}. In the fit, the spectrum $I(\omega)$ is composed of two spectral features residing at different binding energies ($\varepsilon_b$ and $\varepsilon_a$ for the bonding and antibonding bands, respectively), whereby both possess identical single-particle spectral functions $A(\omega)$:
\begin{eqnarray}\label{E1}
I(\omega,T,h\nu) \propto [(M_a(h\nu) A(\omega,\varepsilon_a,T)+ M_b(h\nu)\nonumber\\
\times A(\omega,\varepsilon_b,T)) f(\omega,T)] \otimes R_{\omega} + B(\omega,T),
\end{eqnarray}
\begin{eqnarray}\label{E2}
A(\omega,\varepsilon,T) \propto \frac{|\Sigma''(\omega,T)|}{(\omega - \varepsilon)^2 + \Sigma''(\omega,T)^2},
\end{eqnarray}
where $\Sigma''(\omega,T) = \sqrt{(\alpha \omega)^2 + (\beta T)^2}$ is  the renormalized \cite{Renormalization} imaginary part of the marginal Fermi liquid self-energy with $\alpha = 1.1(1)$, $\beta = 2$ \cite{parameters} and the temperature $T$ in energy units, $f(\omega, T)$ is the Fermi function, $B(\omega)$ is the background which we approximate (assuming it to be $\bf k$-independent) by taking an EDC from the ($\pi/2$, $\pi/2$) point in the empirical form $B(\omega, T) \propto (1 + b\omega^2) f(\omega - \Delta_b(T), T + T_{b})$ with $b$ = 1 eV$^{-2}$, $\Delta_b$(30 K) = 5 meV and $T_{b}$ = 90 K. We are aware that the background could be $h\nu$-dependent, but unless this dependence was very dramatic, the influence on the peak positions and their relative intensities extracted from the fit is marginal. The FWHMs of the resolution Gaussians $R_{\omega}$ were determined for each excitation energy from measurements of the Fermi edge of polycrystalline gold. The components of each spectrum are shown beneath each as thin solid lines: grey lines represent the background, green and blue lines represent the photocurrent from the 'bonding' and 'antibonding' bands, respectively. 

Before going further, we wish to stress three points:

(i) The position of the renormalized band $\varepsilon$, indicated in each fit as a solid vertical line, does {\it not} coincide with maxima in the EDCs, $\omega_m$. This is clearly seen for the 'bonding' peak where $\omega \gg T$ and one can evaluate 
\begin{eqnarray}\label{E3}
\varepsilon = \omega_m \sqrt{1+\alpha^2}
\end{eqnarray} (alternatively, one can write $\varepsilon_{bare} = (1+\lambda)\varepsilon$ for the bare band position, see  \cite{Renormalization}). 

(ii) Although the superconducting gap $\Delta$ can be included into the dispersion, with $\varepsilon$ then becoming $\sqrt{\varepsilon^{2} + \Delta^{2}}$, $\Delta$ cannot exceed the $\varepsilon_a$ value for the antibonding band giving rise to the low energy feature. As will be described below, $\varepsilon_a$ is 11 meV, which is close to $\Delta$ derived in the weak-coupling BCS scheme for given $T_c$. 

(iii) The model upon which the fit is based, and in particular the self-energy function are simple, and represent only one choice from myriad possibilities. In the low energy region the superconductivity has been taken as having no effect on $\Sigma''$, and no coupling to any low-energy bosons was taken into account. Given that the spectral form of feature 'a' depends upon the low-energy properties of the self-energy function, and that the width of this narrow feature is partially determined by the resolution, it is not possible to extract reliable information on the low-energy behavior of $\Sigma''$ from the fit. On the other hand, from the spectral form of feature 'b', one can obtain reliable information on the high-energy behavior of the self-energy function. The fit indicates that $\Sigma''$ at higher binding energies is linear in $\omega$, not only at the antinodal point along the diagonal of the BZ \cite{BorisPRB} but also that this behaviour is a good approximation for higher binding energies at the ($\pi$,0) point. Here we mention that the spectra cannot be fitted with comparable accuracy using a conventional Fermi-liquid self-energy $\Sigma'' \propto \omega^2$.

\begin{figure}[t!]
\includegraphics[width=7cm]{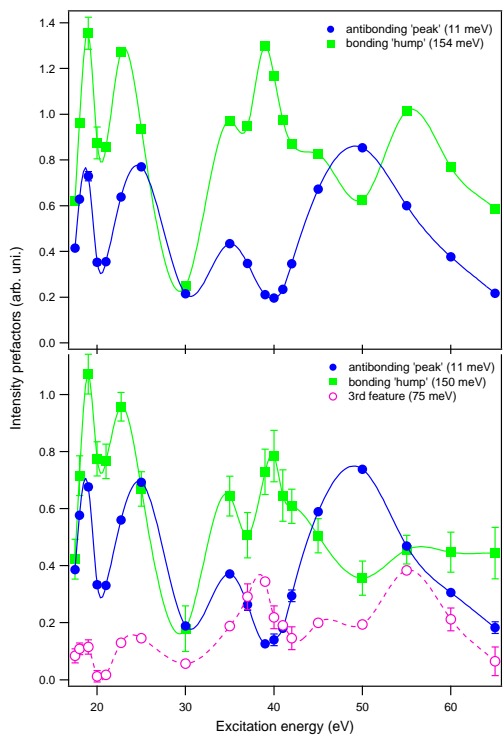}%
\caption{\label{ME} The intensity prefactors $M_a$, $M_b$ and $M_c$ as functions of excitation energy for two-peaks (upper panel) and three-peaks (lower panel) fitting procedure.}
\end{figure}

As a first step, the parameters $\alpha$ = 1.1(1), $\varepsilon_a$ = 11(1) meV, $\varepsilon_b$ = 154(4) meV, $M_a$ and $M_b$ were fixed in the fit using the spectra with $h\nu$=19, 20 and 21 eV, and then all other spectra were fitted using only two free parameters: $M_a$ and $M_b$. These quantities - which we refer to in the following as 'matrix elements' - are represented in the upper panel of Fig.~\ref{ME} as functions of $h\nu$ which clearly shows that the global assignment of the peak and hump in the PDH spectra to two different electronic states is fully justified. However, it should be stated that as there was no $h\nu$ for which either $M_a$ or $M_b$ were exactly zero, we cannot exclude that each individual spectral function has a small, residual PDH-like lineshape.

Turning back to Fig.~\ref{Edep}, one can easily notice that while at some of the excitation energies (18--30, 50 and 65 eV) the two-band fit is very good, at other energies there are small deviations within the binding energy range from 50 to 100 meV. These deviations are photon energy dependent and thus cannot be explained by any kind of an $h\nu$-independent self-energy. This implies the presence of a third feature in these spectra residing at a binding energy of about 75 meV with about 50 meV FWHM. The question then arises as to whether this third feature is related to either the bonding or antibonding band, or whether it is 'extra'. In order to clarify this we carried out fits of all the ($\pi$,0) EDCs now including three non-background components. Such a three-comonent fit perfectly coincides with the experimental data for all photon energies and gives the following positions for the bands: $\varepsilon_a$ = 11(1) meV, $\varepsilon_b$ = 150(4) meV, $\varepsilon_c$ = 75(5) meV, where the c designates the 3rd feature. The $h\nu$-dependence of corresponding matrix elements $M_a$, $M_b$ and $M_c$ are shown in the lower panel of Fig.~\ref{ME}. 

The fact that the $M_b(h\nu)$ and $M_a(h\nu)$ functions exhibit much the same form for both the 2 or 3-component fits shows that this result is robust with respect to the introduction of the 3rd peak. Consequently, Fig.~\ref{ME} provides two pieces of evidence that exactly these two features are originated from the splitted CuO band: (1) the $h\nu$-average values of $M_b$ and $M_a$ are comparable ($\langle M_b \rangle / \langle M_a \rangle$ = 1.2) as could well be expected from a pair of bands of the same atomic character split by the c-axis bilayer coupling while the average value for $M_c$ is much smaller ($\langle M_b \rangle / \langle M_c \rangle$ = 2.9); (2) the $h\nu$-dependencies of $M_b$ and $M_a$ are in a good agreement with recent calculations \cite{LindroosXXX}. Further support for this assignment comes from the fact that the energetic separation between the renormalized bands' positions is about 140 meV, which leads to a difference between the peak maxima in the EDCs of 65-85 meV, in keeping with the normal state splitting observed recently in \cite{ChuangXXX}, or to a difference in the bare band positions of about 300 meV (assuming $\lambda \approx 1$ \cite{LanzaraNature} in Eq.\ref{E3} which we also obtain from a Kramers-Kr\"onig transform of $\Sigma''(\omega)$) which is consistent with the results of the tight-binding calculations \cite{Liechtenstein}.

As an additional evidence, we show in Fig.~\ref{GM} the series of EDCs along the (0,0)-($\pi$,0) direction in the BZ. That was the second big cornerstone in PDH physics - the dispersionless nature of the 'superconducting' peak in this direction \cite{NormanPRL97}. We choose the overdoped modulation-free (Pb,Bi)-2212 at $h\nu$ = 21.2 eV (at which the 3rd feature is weak) above and below $T_c$ to demonstrate the conventional dispersion of the peak-forming band. It turns out that the temperature, like polarization or variable excitation energy, can also be used to control the relative intensities of the the bonding and antibonding components. At 300 K the actual amplitude of the antibonding peak, which is much closer to the Fermi level than the bonding one and therefore is more strongly influenced by temperature (see Eq.\ref{E2}), is strongly cut by the Fermi function and has twice smaller matrix elements (see Fig.~\ref{ME}), is noticeably less than the amplitude of the bonding peak which mainly determines the position of the room temperature EDC maxima. The substitution of the given temperatures into Eq.1 gives a good coincidence with the experimental spectra. Thus, it is the bonding peak which mainly determines the position of the room temperature EDC maxima, whereas the maxima of low temperature spectra probe the position of the antibonding band. 

\begin{figure}[t!]
\includegraphics[width=8.47cm]{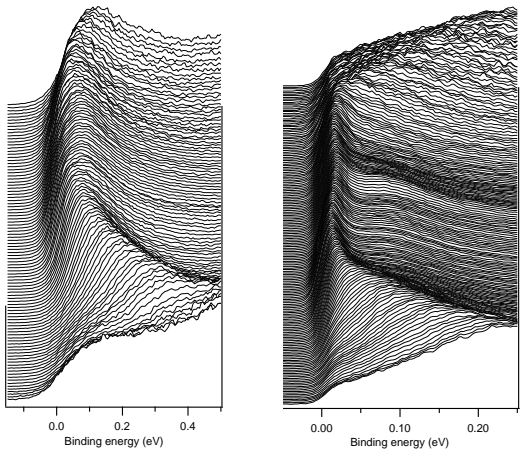}%
\caption{\label{GM} The series of EDCs, self-normalized to its maximum, along (0,0)-($\pi$,0) direction in BZ for 300K (left panel) and 39K (right panel).}
\label{GM}
\end{figure}

In order to discuss the nature of the 3rd feature one can consider the following possibilities: (i) surface states, (ii) BiO pockets, and (iii) intensity from a 'shadow band'. We can exclude a macroscopic inhomogenity of the crystals on the grounds of their perfect low energy electron diffractograms and sharp superconducting transitions (both measured for every cleave). While we do not exclude completely the first two possibilities (although it is widely belived that there are no conducting states at the top BiO layer, the intensity of feature c in these data is rather low and thus it is conceivable that these states are not easily seen in tunneling experiments \cite{BiO}), the 'shadow band' would seem to be a more likely candidate. Up to now, there is no unambiguous proof neither for its structural nor antiferromagnetic nature. Then, if one would accept that the shadow FS is at least partially connected with an antiferromagnetic phase (possibly originating from microscopic inhomogeneities \cite{Pan}), then it is quite natural to expect that its band at the ($\pi$,0) point will not coincide with either the bonding nor the andibonding CuO bands of the main phase. This kind of picture could also offer an alternative explaination of the PDH-like spectra seen in ARPES experiments in the nodal direction \cite{LanzaraNature}. 

Finally, we note that the interpretation of the PDH lineshape in terms of different underlying electronic bands is {\it not} restricted to the OD case. We have also observed a similar photon energy dependence of the PDH features for an underdoped sample. The detailed analysis of these data will be presented in a forthcoming paper.

To sum up, we have presented a detailed investigation of the superconducting state ($\pi$,0)-photoemission spectra of overdoped, modulation-free Pb-Bi2212 single crystals. We demonstrate that the PDH lineshape is strongly dependent on the excitation energy, which is practically irreconcilable with models in which the peak, dip and hump are considered to stem from a single band spectral function. The lineshape of the spectra can be quantitatively well reproduced by the superposition of spectral features described by essentially identical single-particle spectral functions but residing at different binding energies: the hump is mainly formed by the bonding band and the peak originates from the antibonding band. Models for superconductivity in cuprates based on a single band PDH structure should be reconsidered.

We acknowledge stimulating discussions with S.-L. Drechsler, A. N. Yaresko and H. Eschrig. We are grateful to B. Keimer for suplying the pure Bi2212 sample, to S. Cramm for the technical assistance and to the DFG (Graduiertenkolleg "Struktur- und Korrelationseffekte in Festk\"orpern" der TU-Dresden) and to the Fonds National Suisse de la Recherche Scientifique for support.


\begin{references}
\bibitem{Renormalization} The real part of the self-energy is included into its imaginary part and band position $\varepsilon$ as a result of renormalization. If, for simplicity, one assumes $\Sigma'(\omega) = -\lambda\omega + o(\omega^3)$ ($\lambda > 0, \omega > 0$) than such a renormalization can be done by simple scaling procedure: $\varepsilon/(1 + \lambda) \rightarrow \varepsilon$, $\Sigma''/(1 + \lambda) \rightarrow \Sigma''$, $(1 + \lambda) A \rightarrow A$.
\bibitem{parameters} These parameters are similar to the values derived from a similar analysis of the EDCs recorded along the $\Gamma$-($\pi$,$\pi$), or nodal direction \cite{BorisPRB}.
\end{references}
\end{document}